\documentclass{elsart}
\usepackage{amstext,amsfonts,amssymb,amsbsy,array,epic,eepic,graphicx, amsmath,epstopdf}
\usepackage{stmaryrd}

\newcommand{\mcE}{\mathfrak{F}}

\newcommand{\bff}{{\boldsymbol f}}

\newcommand{\bfK}{{\boldsymbol K}}
\newcommand{\bfS}{{\boldsymbol S}}
\newcommand{\bfg}{{\boldsymbol g}}
\newcommand{\bfh}{{\boldsymbol h}}
\newcommand{\bfx}{{\boldsymbol x}}

\newcommand{\bfn}{{\boldsymbol n}}
\newcommand{\bft}{{\boldsymbol t}}

\newcommand{\bfI}{{\boldsymbol I}}

\newcommand{\bfW}{{\boldsymbol W}}

\newcommand{\bfu}{{\boldsymbol u}}
\newcommand{\bfeps}{{\boldsymbol\varepsilon}}
\newcommand{\bfv}{{\boldsymbol v}}

\newcommand{\bfsig}{{\boldsymbol\sigma}}

\newcommand{\jump}[1]{\left\llbracket #1\right\rrbracket}

\newcommand{\triangulation}{\mathcal{T}^h}
\def\eref#1{{\rm (\ref{#1})}}                                    
\begin{document}
\begin{frontmatter}
\title{A discontinuous Galerkin method for cohesive zone modelling}
\author{Peter Hansbo}\and
\author{Kent Salomonsson}
\address{Department of Mechanical Engineering, J\"onk\"oping University, S-551 11 J\"onk\"oping,
Sweden} 
%
\maketitle
\begin{abstract}
We propose a discontinuous finite element method for small strain elasticity allowing for cohesive zone modeling. 
The method yields a seamless transition between the discontinuous Galerkin method and classical cohesive zone modeling.
Some relevant numerical examples 
are presented.
\end{abstract}
\end{frontmatter}
\section{Introduction}
\label{intro}
In this paper we develop a discontinuous finite element 
method for cohesive zone modeling using the approach first suggested by Hansbo and Hansbo \cite{HaHa04}. 
Unlike in
the standard pre--failure treatment of cohesive zones, which consists of
tying the meshes together using a penalty approach, we use a combination of Nitsche's method and the cohesive law
governing the interelement stiffness, thus allwoing the same discretization method in both pre--failure and post--failure 
regimes.
This means that the method is consistent with the original differential equation and 
no large penalty parameters are required for accurate solutions even in the pre--failure regime. 
The approach was implemented for cohesive cracks
by Heintz and Hansbo \cite{HaHe06} in an XFEM setting, but here we consider a discontinuous Galerkin method allowing
for discontinuities appearing only between elements. 

An approach similar to ours has been suggested by Mergheim, Kuhl, and Steinmann \cite{MeKuSt04}, 
an later used by Pretchel et al. \cite{PrRoJaHaLeStSt11} and Wu et al. \cite{WuTjBeMaJeNo13}. The method of \cite{MeKuSt04} however uses a different
blending of Nitsche's method and cohesive zones. There a discontinuous Galerkin method
is used only in the pre--failure regime and a switch to a standard cohesive zone approximation is performed at a given traction threshold.
To ensure a continuous transition between the discretization methods, a matching of discrete 
tractions between the two cases is performed. This matching is cumbersome in a more general situation of 
nonmatching meshes across the cohesive zone. In this paper we avoid this switch and a more generally applicable method results.

An outline of the remainder of the paper is as follows. In Section 2 we define our discrete method in a linear setting; in Section 3 we discuss and motivate the cohesive law that we favour and the resulting secant compliance we use in our numerical solution process; and in Section 4 we give some numerical examples of our approach. Finally, in Section 5, we give some concluding remarks.

\section{The model problem and discretization method}

\subsection{Linear elasticity with a single cohesive interface}

W consider first an elasticity problem
in $n_{\text{sd}}=2$ or $3$ dimensions with a smooth boundary $\Gamma$ dividing $\Omega$ into two parts $\Omega_1$ and $\Omega_2$.
The displacement$\bfu = \left[
u_i\right]_{i=1}^{n_{\text{sd}}}$ has restrictions to the different domains $\bfu_i = \bfu\vert_{\Omega_i}$, and we denote
by $\jump{\bfu} = \bfu_1\vert_\Gamma -\bfu_2\vert_\Gamma$ and let $\bfn$ denote the outward normal vector to $\partial\Omega$ and to $\Omega_1$ on $\Gamma$. Then, a linear elasticity problem with cohesive layer $\Gamma$ can be written: Find $\bfu$ and and the symmetric stress tensor $\bfsig =
\left[\sigma_{ij}\right]_{i,j=1}^{n_{\text{sd}}}$ such that
\begin{equation}\label{diffelasti}
\begin{array}{r} 
  \bfsig = \lambda ~\nabla\cdot\bfu\bfI 
   + 2 \mu \bfeps(\bfu)\quad \text{in $\Omega_1\cup\Omega_2$},\\ 
-\nabla\cdot\bfsig = \bff \quad \text{in $\Omega_1\cup\Omega_2$}, \\ 
\bfu = \bfg \quad \text{on $\partial\Omega_{\text{D}}$}, \\ 
\bfsig\cdot\bfn = \bfh  \quad \text{on $\partial\Omega_{\text{N}}$}\\ 
\jump{\bfsig\cdot\bfn}  = 0\quad \text{on $\Gamma$}\\
\jump{\bfu}  =  -\bfK \bfsig\cdot\bfn \quad \text{on $\Gamma$}
\end{array}
\end{equation}
Here $\lambda$ and $\mu$ are positive constants called the 
Lam\'{e} constants, satisfying $0 <\mu_1 < \mu < \mu_2$ and 
$0< \lambda < \infty,$ and 
$\bfeps\left(\bfu\right) = \left[\varepsilon_{ij}(\bfu)\right]_{i,j=1}^{n_{\text{sd}}}$ 
is the strain tensor with components
\[ \varepsilon_{ij}(\bfu) = \frac{1}{2}\left( \frac{\partial
  u_i}{\partial x_j}+\frac{\partial u_j}{\partial x_i}\right).
\] Furthermore,
$\nabla\cdot\bfsig = \left[\sum_{j=1}^{n_{\text{sd}}}\partial
  \sigma_{ij}/\partial x_j\right]_{i=1}^{n_{\text{sd}}}$,
$\bfI = \left[\delta_{ij}\right]_{i,j=1}^{n_{\text{sd}}}$ with $\delta_{ij} =1$
if $i=j$ and $\delta_{ij}= 0$ if $i\neq j$, $\bff$ and $\bfh$ 
are given loads, $\bfg$ is a given boundary displacement, and $\bfn$ is
the outward unit normal to $\partial\Omega$. 
Finally, $\bfK$ is a symmetric positive semi--definite flexibility matrix (constitutive law on $F$). For example, with isotropic elasticity on $F$ we have that
\[
\bfK=\alpha\,\bfI + (\beta-\alpha)\bfn\otimes\bfn,\quad\text{or}\quad K_{ij}=\alpha\delta_{ij}+(\beta-\alpha)n_in_j, 
\]
where $\otimes$ denotes outer product,
with $\alpha \geq 0$ and $\beta\geq 0$ denoting the complicancy in the direction tangential and normal to $F$, respectively, cf. \cite{HaHa04}. (In this paper, a more general compliance, with cross coupling between normal and tangential directions will be considered.)

\subsection{A discontinuous Galerkin method for linear cohesive zones}

Consider a subdivision of $\Omega$ into a geometrically conforming finite element partitioning
$ \triangulation =\{ T\}$ of $\Omega$. Let
\[
P^k(T) =\{\text{$\bfv$: each component of $\bfv$ is a polynomial of degree $\leq k$ on
$T$}\},
\]
\[
\bfW^h = \{\bfv\in [L^2(\Omega)]^{n_{\text{sd}}}:~\bfv\vert_T\in [P^k(T)]^{n_{\text{sd}}}~~ \forall T \in \triangulation \}.
\]
We also introduce the set
of element faces in the mesh, $\mcE =\{ F \}$, and we split $\mcE$
into three disjoint subsets
\[
  \mcE = \mcE_I \cup \mcE_D\cup \mcE_N,
\]
where $\mcE_I$ is the set of faces in the interior of $\Omega$ and
$\mcE_D$ and $\mcE_N$ are the sets of faces on the Dirichlet and Neumann part of the boundary, respectively. Further, with each
face we associate a fixed unit normal $\bfn$  such that for faces
on the boundary $\bfn$ is the exterior unit normal. 
We denote the
jump of a function $\bfv\in\bfW^h$ at an internal face $F\in \mcE_I$ by $ \jump{\bfv}
 = \bfv^+-\bfv^-$,  and $
\jump{\bfv} = \bfv^+$ for $F \in \mcE_D$, 
and the average
$\langle  \bfv\rangle = (\bfv^+ + \bfv^-)/2$ for $F\in \mcE_I$,
and $\langle \bfv\rangle = \bfv^+$ for $F \in \mcE_D$, 
where
$v^{\pm} = \lim_{\epsilon\downarrow 0} \bfv(\bfx\mp
\epsilon\,\bfn)$ with $\bfx\in F$.

For the modelling of cohesive interfaces, we here assume that the solution may be discontinuous across each element face $F$,
and thus the role of $\Gamma$ in (\ref{diffelasti}) is now taken by \emph{all}\/ element faces.

The DG method can then be formulated as follows: Seek $\bfu^h\in \bfW^h$ such that
\begin{equation}\label{niteq}
a_h(\bfu^h,\bfv)=L_h(\bfv) \quad \text{for all $\bfv \in \bfW^h$}.
\end{equation}
The bilinear form is defined by
\begin{equation}\label{nitsche_form}\begin{array}{>{\displaystyle}l}
a_h(\bfu^h,\bfv)=\sum_{T\in\triangulation}\int_T
\bfsig(\bfu^h):\bfeps(\bfv)\,dx 
\\ 
 \, -\sum_{F\in \mcE_I\cup\mcE_D}\int_{F}\left<\bfsig(\bfu^h)\cdot\bfn\right>\cdot\left(\jump{\bfv} + \bfK\left<\bfsig(\bfv)\cdot\bfn\right>\right)
\, ds \\ 
 \,   -\sum_{F\in \mcE_I\cup\mcE_D}\int_{F}\left<\bfsig(\bfv)\cdot\bfn\right>\cdot\left(\jump{\bfu^h} + \bfK\left<\bfsig(\bfu^h)\cdot\bfn\right>\right)
\, ds \\ 
 \,   +\sum_{F\in \mcE_I\cup\mcE_D}\int_{F}\left<\bfsig(\bfv)\cdot\bfn\right>\cdot\left(\bfK\left<\bfsig(\bfu^h)\cdot\bfn\right>\right)
\, ds \\ 
 \,   + \sum_{F\in \mcE_I\cup\mcE_D}\int_{F}\left(\bfS_h\left(\jump{\bfu^h} + \bfK\left<\bfsig(\bfu^h)\cdot\bfn\right>\right)\right)\cdot\left(\jump{\bfv} + \bfK\left<\bfsig(\bfv)\cdot\bfn\right>\right)\, ds,
\end{array}\end{equation}
and the linear functional by
\begin{eqnarray}
L_h(\bfv) &=&\int_\Omega\bff\cdot\bfv \, dx
+\sum_{F\in\mcE_N}\int_{F}\bfh\cdot\bfv \,ds -\sum_{F\in\mcE_D}\int_{F}\bfsig(\bfv)\cdot\bfn \cdot \bfg \,ds 
\\ \nonumber
&& \quad + \sum_{F\in\mcE_D}\int_{F}\left(
\bfS_h\bfg\right)\cdot \left(\bfv+\bfK\bfsig(\bfv)\cdot\bfn\right)\, ds .
\end{eqnarray}
 Here  $\bfS_h$ is a  matrix which depends on the interface conditions of the problem, the local meshsize, and a penalty parameter $\gamma : = (2\mu + 3\lambda )\gamma_0$, where $\gamma_0$ is a dimensionless number which has to be large enough for the method to be stable. The
 stability of the method increases with increasing flexibility, so the choice of $\gamma_0$ needed for 
 stability in the case of zero flexibility can be used in all other cases (numerical values for $\gamma_0$ can be found, e.g.,  in \cite{HaLa02}). More precisely, on a face $F$ with diameter $h_F$,
\begin{equation} \label{sh}
\bfS_h|_F =\left(\frac{h_F}{\gamma}\bfI  + \bfK\right)^{-1}
\end{equation}
On each face $F$, the mesh parameter $h_F$ is defined by 
\begin{equation}
h_F =
\begin{cases}
 \bigl( \text{meas}(T^+) + \text{meas}(T^-) \bigl) / 2 \, \text{meas}(F) 
        &  \text{ for $F \subset \partial T^+ \cap \partial T^-$},
\\
 \text{meas}(T)/ \text{meas}(F)  & \text{ for $F \subset \partial T \cap \partial\Omega_D$}. 
\end{cases}
\end{equation}

We note that as the flexibility goes to zero, we approach a standard discontinuous Galerkin method for elasticity. Looking instead at the limit case of $h\rightarrow 0$ (assuming $\bfK$ is invertible) we retrieve a standard formulation for cohesive laws where the only term contributing to the stiffness matrix from the interfaces is the interface stiffness term
\[
\sum_{F\in \mcE_I\cup\mcE_D}\int_{F}\left(\bfK^{-1}\jump{\bfu^h}\right)\cdot\jump{\bfv} \, ds .
\]
The proposed method thus seamlessly blends discontinuous Galerkin with standard FEM for cohesive interfaces.

By use of Green's formula, we readily establish that the method \eref{niteq} is consistent in the sense that
\begin{equation}\label{galort}
a_h(\bfu-\bfu^h,\bfv) = 0
\end{equation}
for all $\bfv\in \bfW^h$ and for $\bfu^h$ sufficiently regular, which is key to retrieving optimal accuracy of the method.
Stability follows from the analysis in \cite{Br04,HaHa04,HaLa02}. We also mention the work of Juntunen and Stenberg \cite{JuST09}, where
an analysis of this approach for handling general boundary conditions for Poisson's equation is given.

\section{Cohesive law implementation\label{implement}}

We are now interested in the case when the interface compliance depends on the jump of the solution, $\bfK= \bfK(\jump{\bfu})$.
In the numerical solution of the cohesive zone FE model, we replace this compliance by a corresponding secant compliance as follows.

The compliance between the elements can be defined a priori by cohesive zone models. Frequently, cohesive zone models  that  are easy to implement are chosen to model the initiation of cracks. These models are often un-coupled, meaning that there is no relationship between the normal and tangential stresses other than the fracture criterion. However, it is reasonable to imagine that the tangetial stiffness is effected by the reduction of the normal stiffness. Thus, in order to couple the cohesive behavior of the interfaces in mixed mode, we choose to derive the traction-separation laws from a weighted energy release rate surface, cf. \cite{SaAn10,AnSaTh14}.\par

We denote the energy release rates in pure normal and pure tangetial directions $\mathnormal{\Gamma}_\mathrm{I}\left(u_n\right)$ and $\mathnormal{\Gamma}_\mathrm{II}\left(u_t\right)$, respectively, where, for convenience, the normal and tangetial jumps in displacement are denoted by $u_n= \jump{\bfu}\cdot\bfn$ and $u_t= \jump{\bfu}\cdot\bft$, where $\bft$ is the tangent vector to the given face, such that $\bfn$ and $\bft$ constitute a right-handed ON system. The energy release rates are obtained from the interface traction on $\mcE_I$ according to
 
\begin{equation}
        \mathnormal{\Gamma}_\mathrm{I}\left(u_n\right)=\mathnormal{\Gamma}_\mathrm{I}\left(u_n,0\right)= \int\limits_0^{u_n}\bfsig\left(\tilde{u_n},0\right)\cdot\bfn\,\mathrm{d}\tilde{u_n},
       \label{Ji11}
\end{equation}
 
\begin{equation}
        \mathnormal{\Gamma}_\mathrm{II}\left(u_t\right)=\mathnormal{\Gamma}_\mathrm{I}\left(0,u_t\right)= \int\limits_0^{u_t}\bfsig\left(0,\tilde{u_t}\right)\cdot\bft\,\mathrm{d}\tilde{u_t}.
       \label{Ji22}
\end{equation}
 
By use of a polar coordinate system, a dimensionless effective separation $\lambda$ can be defined together with an angle $\varphi$ that determines the mode mix. The mode mix and the effective separation are defined as
 
\begin{equation}
       \varphi=\arctan\left(\frac{u_\mathrm{nc}u_n}{u_\mathrm{tc}u_t}\right),
       \label{phi}
\end{equation}

\begin{equation}
       \lambda=\sqrt{\left(\frac{u_n}{u_\mathrm{nc}}\right)^{2}+\left(\frac{u_t}{u_\mathrm{tc}}\right)^{2}}
       \label{lam}.
\end{equation}

\begin{figure}[h]
	\centering
		\includegraphics[scale=0.8]{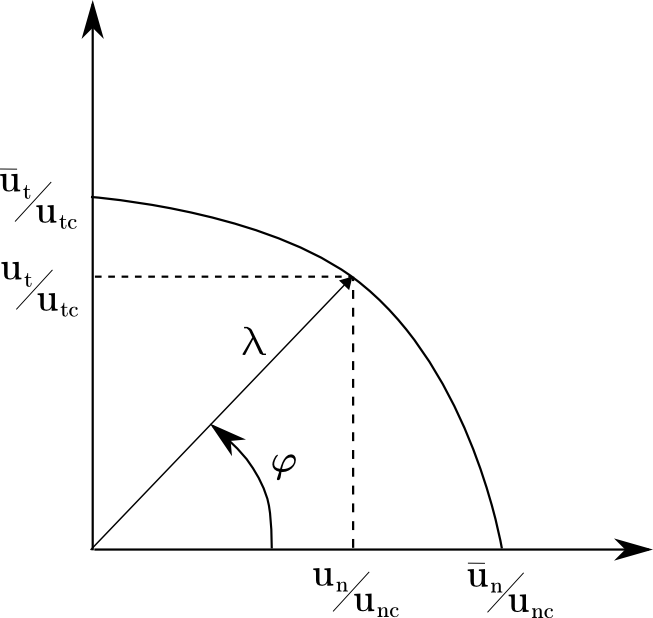}
	\caption{Illustration of effective separation and mode-mixity}
	\label{fig:Fig3}
\end{figure}

Here, $u_\mathrm{nc}$ and $u_\mathrm{tc}$ are the critical normal and tangential separations in pure modes. The normal and tangential separations, $u_n/u_\mathrm{nc}$  and $u_t/u_\mathrm{tc}$ , are defined as the projections of the effective separation on each respective pure mode axis, cf. Fig. \ref{fig:Fig3}. 

It then follows that $u_n$ and $u_t$ are given by

\begin{equation}
       u_n=\lambda u_\mathrm{nc} \cos\left(\varphi\right)
       \label{w},
\end{equation}

\begin{equation}
       u_t=\lambda u_\mathrm{tc} \sin\left(\varphi\right)
       \label{w}.
\end{equation}

\begin{figure}[htbp]
	\centering
		\includegraphics{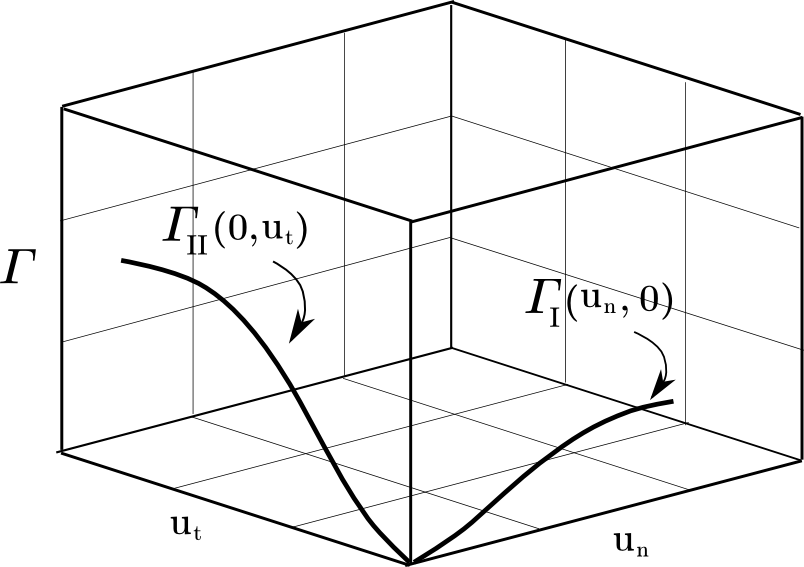}
	\caption{Schematic illustration of the fitted energy release rate curves in normal and tangential directions.}
	\label{fig:Fig4}
\end{figure}

In order to obtain the complete contributions of the energy release rates in each pure mode and not the projections, some additional definitions, $\overline{u}_n$ and $\overline{u}_t$ are introduced. For example, we may choose to define $\overline{u}_n:=\lambda u_\mathrm{nc}$ and $\overline{u}_t:=\lambda u_\mathrm{tc}$. As a first step in the development of the cohesive law, two independent functions are fitted to experimentally measured energy release rate curves, see $\mathnormal{\Gamma}_\mathrm{I}$ and $\mathnormal{\Gamma}_\mathrm{II}$ in Fig.\ref{fig:Fig4}.

\begin{figure}[htbp]
	\centering
		\includegraphics[scale=0.6]{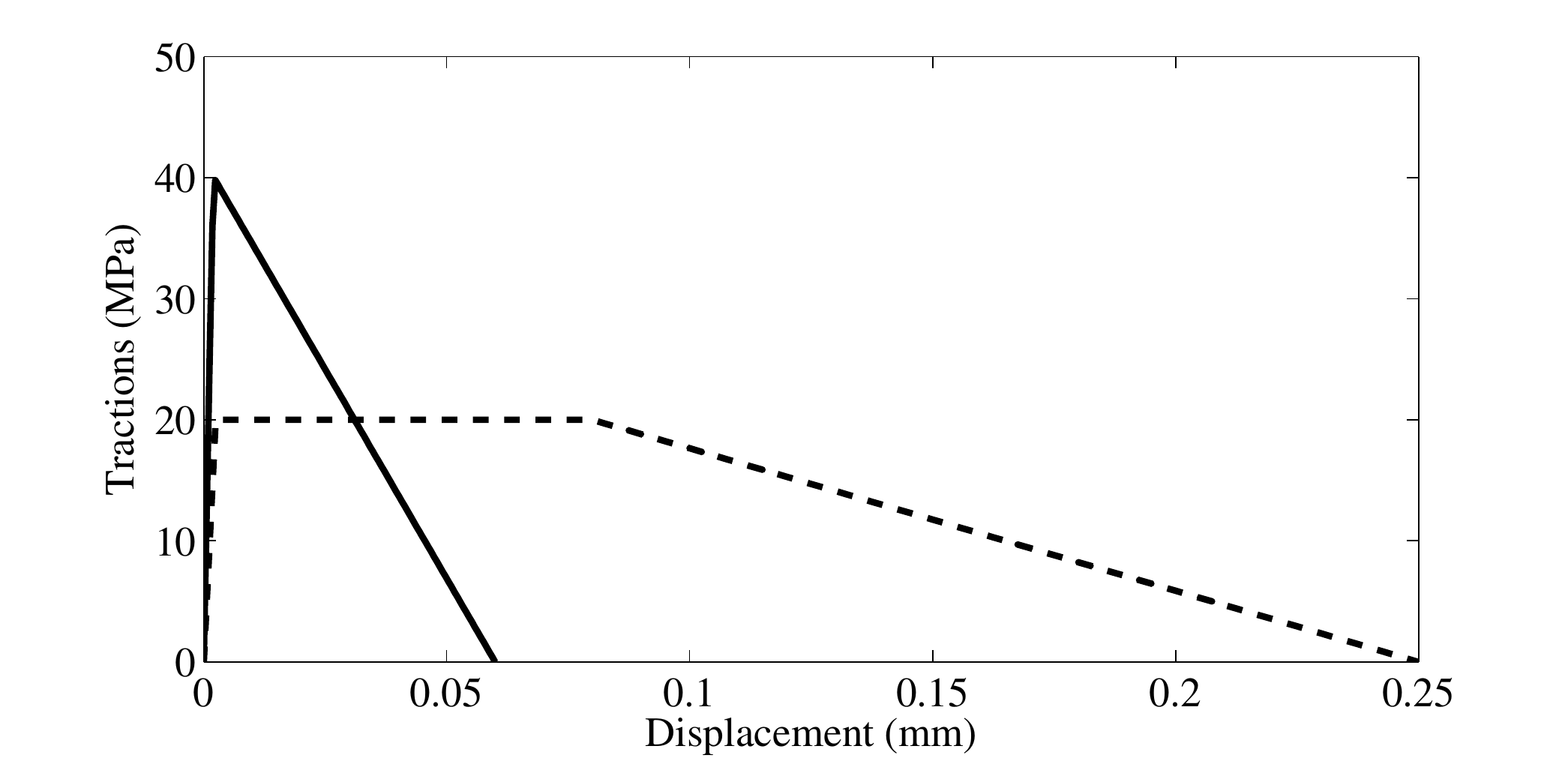}
	\caption{Schematic illustration of normal (solid) and tangential (dashed) traction-separation curves.}
	\label{fig:Fig5}
\end{figure}

The shapes of the traction-separation curves in each pure mode, respectively, are obtained by differentiating the energy release rates in each pure mode with respect to each pure mode relative separation, $u_n$ and $u_t$. From these curves, laws are chosen that captures the most essential features of the curves. Figure \ref{fig:Fig5} shows two idealized schematic curves. 

\begin{figure}[htbp]
	\centering
		\includegraphics[scale=0.6]{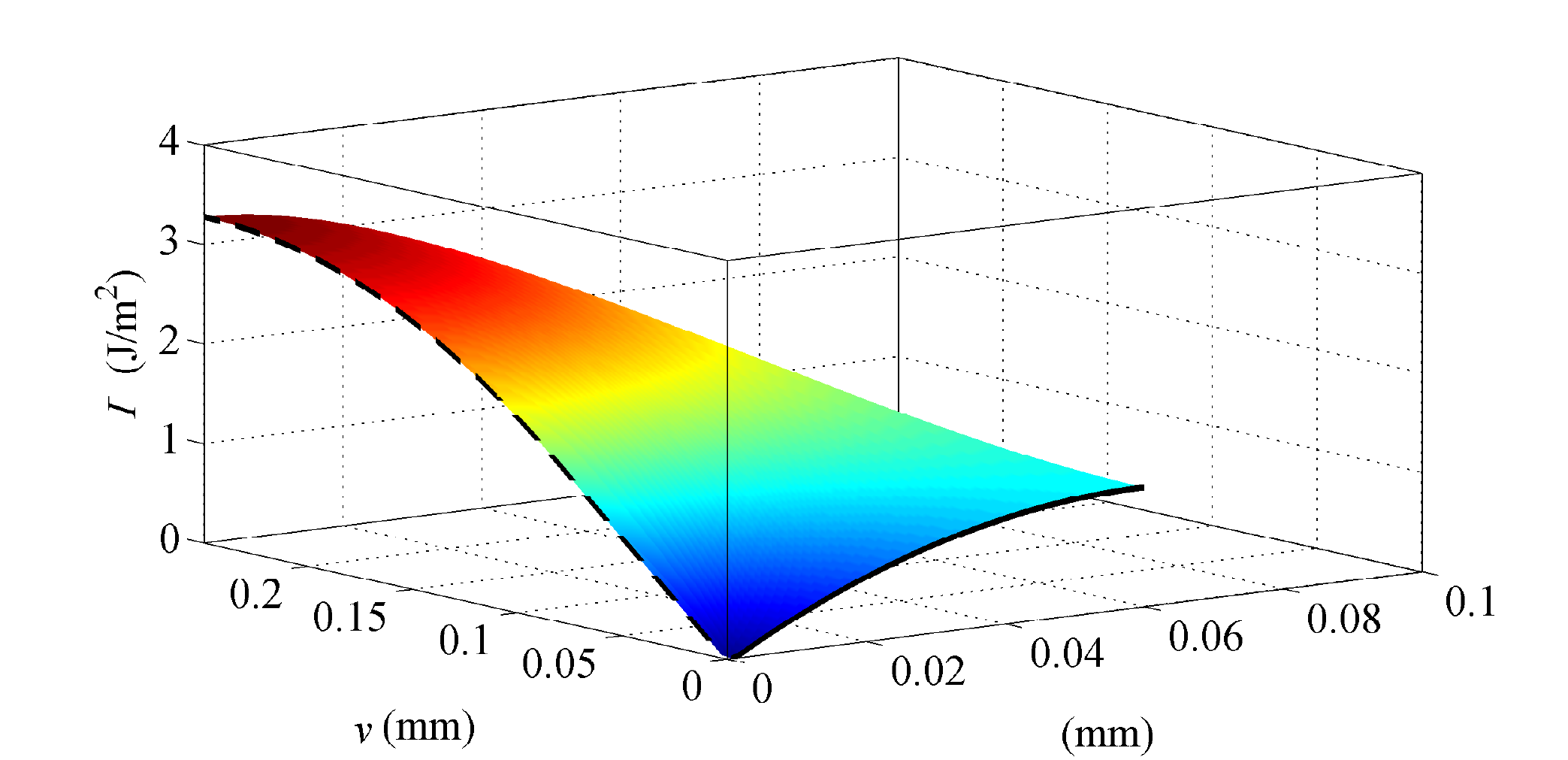}
	\caption{Weighted potential surface.}
	\label{fig:Fig6}
\end{figure}

In order to capture the behavior of the cohesive law in mixed mode, the two energy release rate curves in Fig. \ref{fig:Fig4} are combined to yield a surface where the axes are total energy release rate, $\mathnormal{\Gamma}$, relative normal and relative tangential separations, $u_n$ and $u_t$ respectively, see Fig. \ref{fig:Fig6}. \par

The surface representing the weighted energy release rate, $\mathnormal{\Gamma}\left(\lambda,\varphi\right)$ is generated by a weighted sum of the experimentally determined energy release rates in pure normal, $\mathnormal{\Gamma}_\mathrm{I}$, and pure tangential, $\mathnormal{\Gamma}_\mathrm{II}$, directions according to 

\begin{equation}
	   \mathnormal{\Gamma}_\mathrm{I}\left(\lambda,\varphi\right)=f\left(\varphi\right)\mathnormal{\Gamma}_\mathrm{I}\left(\lambda\right){u}^{2}_\mathrm{nc} + \left(1-f\left(\varphi\right)\right)\mathnormal{\Gamma}_\mathrm{II}\left(\lambda\right)u^{2}_\mathrm{tc}
	   \label{J}
\end{equation}

where $f\left(\varphi\right)$ is the weight function. \par 

The stresses for any given mode mix are given by partial differentiation of $\mathnormal{\Gamma}$ with respect to each relative separation, $u_n$ and $u_t$, respectively.

\begin{equation}
	   \bfsig\cdot\bfn=\frac{\partial\mathnormal{\Gamma}}{\partial u_n}=\frac{\partial\mathnormal{\Gamma}}{\partial \lambda} \frac{\partial\lambda}{\partial u_n}+\frac{\partial\mathnormal{\Gamma}}{\partial \varphi} \frac{\partial\varphi}{\partial u_n}
	   \label{sig},
\end{equation}

\begin{equation}
	   \bfsig\cdot\bft=\frac{\partial\mathnormal{\Gamma}}{\partial u_t}=\frac{\partial\mathnormal{\Gamma}}{\partial \lambda} \frac{\partial\lambda}{\partial u_t}+\frac{\partial\mathnormal{\Gamma}}{\partial \varphi} \frac{\partial\varphi}{\partial u_t}
	   \label{tau}
\end{equation}

The secant compliance is computed as follows. We first establish the secant stiffness matrix $\bfS_T$ as
\[
\bfS_T = \frac{\bfsig\cdot\bfn}{u_n}\bfn\otimes\bfn + \frac{\bfsig\cdot\bft}{u_t}\bft\otimes\bft+ \frac{\bfsig\cdot\bft}{u_n}\bfn\otimes\bft +\frac{\bfsig\cdot\bfn}{u_t}\bft\otimes\bfn ,
\]
followed by computing the secant compliance as $\bfK_T = \bfS_T^{-1}$.
%
The interface stiffness, $\bfS_h|_F$, is then given by 

\begin{equation} \label{sh}
\bfS_h|_F =\left(\frac{h_F}{\gamma}\bfI  + \bfK_T\right)^{-1}
\end{equation}

\section{Numerical example}
\begin{figure}[htbp]
	\centering
		\includegraphics[scale=0.65]{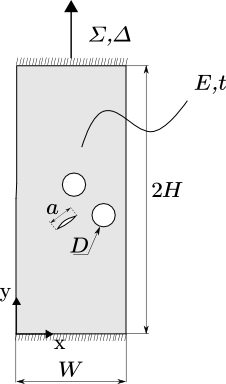}
	\caption{Dimensions of the Single Edge Notched specimen.}
	\label{fig:SIMspec}
\end{figure}

A specimen with two inclusions and an initial crack, see Fig. \ref{fig:SIMspec}, is used as a simple example to show the applicability of the modeling technique. The dimensions of the specimen are given by; $W=H=1.00$ mm, $D=0.20$ mm, $a=0.20$ mm. The lower right inclusion is located at center coordinate $(0.75, 1.00)$ mm and the top left inclusion is located at $(0.45,1.10)$ mm. The crack is located at center coordinates $(0.40,0.90)$ mm and it is inclined at an angle of 33$^\circ$ to the horizontal axis. The boundary conditions for the specimen are set to be clamped on the bottom edge, i.e. $u_x(x,0) = u_y(x,0) = 0$. The top boundary is constrained horizontally $u_x(x,\it{2H})=0$ and the displacement is controlled vertically $u_y(x,\it{2H})=\Delta$, see Fig. \ref{fig:SIMspec}. Two different set-ups are modeled for comparison. The first is a specimen where the inclusions have the same material properties as the rest of the specimen with elastic material properties;  $E=10$ MPa and $\nu=0.45$. The second is a specimen where the Young's modulus of the inclusions is 100 times greater than in the rest of the specimen. The maximum cohesive strengths are set to 1 MPa and the maximum critical separations are set to 0.02 mm in the cohesive sawtooth model giving a fracture energy of $0.1$ $\rm{J/{mm}^2}$. Note that these properties are the same for both set-ups. \par 

One of the major issues with this type of modeling is mesh dependency. However, if a large number of elements is used the mesh dependency is obviously reduced. Furthermore, the compliance between all continuum elements introduce numerical issues which can be reduced by an increase of the elastic stiffness of the cohesive zone model sufficiently to minimize the compliance.\par

In the present model, the compliance is allowed to be initially zero and then gradually increase as the load is increased. Damage initiation, and essentially crack propagation, is enabled by a decrease of the stiffness according to \eref{sh} where the interfaces, as stated in the definition of the method, are given as the boundaries between all the continuum elements 
(this is of course not a requirement, as a mix of continuous and discontinuous methods is also possible). Thus, cracks are free to form, nucleate and propagate along the continuum element boundaries by Nitsche's method instead of the standard approach of using cohesive elements, as in, e.g., \cite{SaAn08,XuNe94}.  \par

\begin{figure}[htbp]
	\centering
		\includegraphics[scale=0.28]{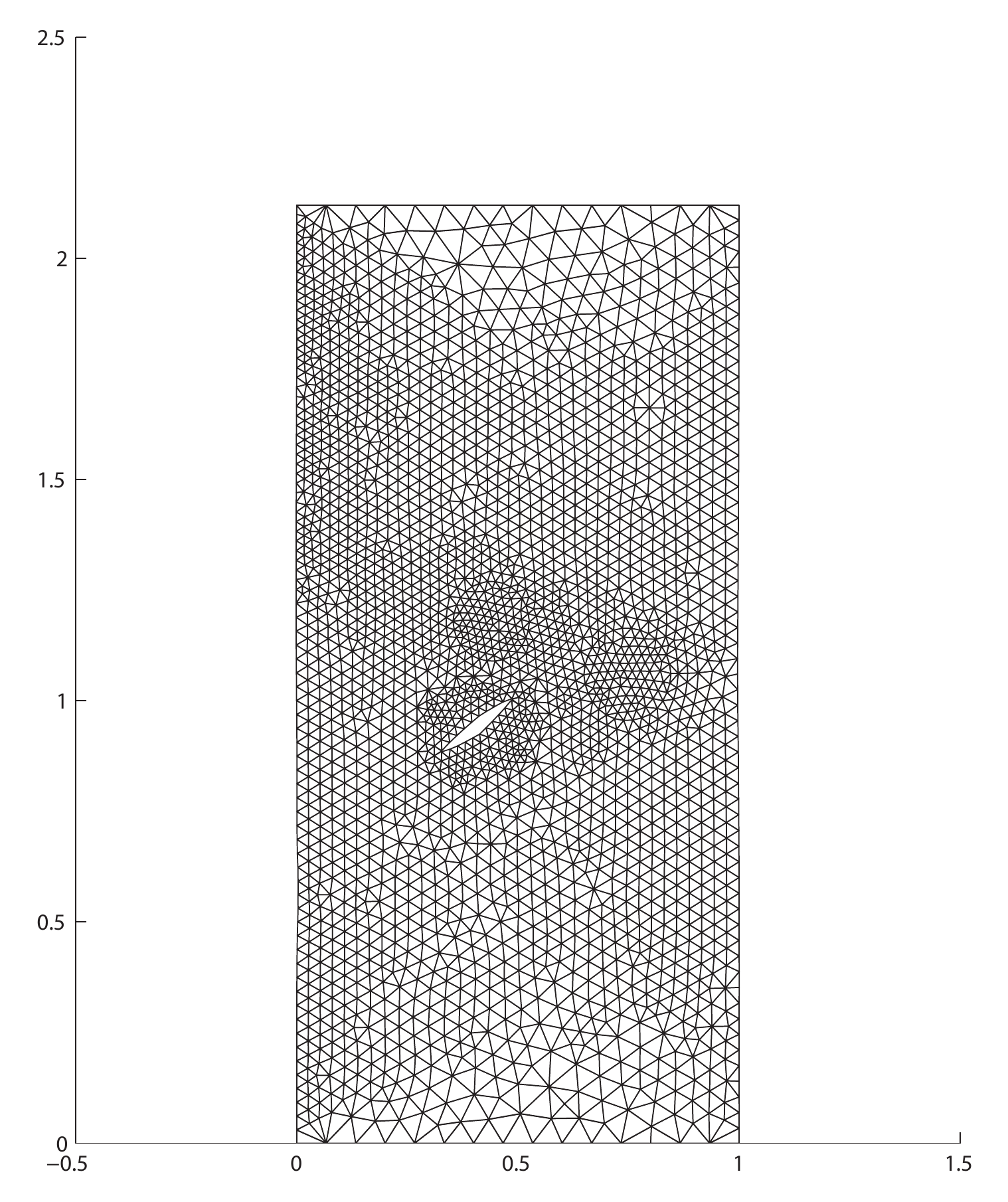}
		\includegraphics[scale=0.28]{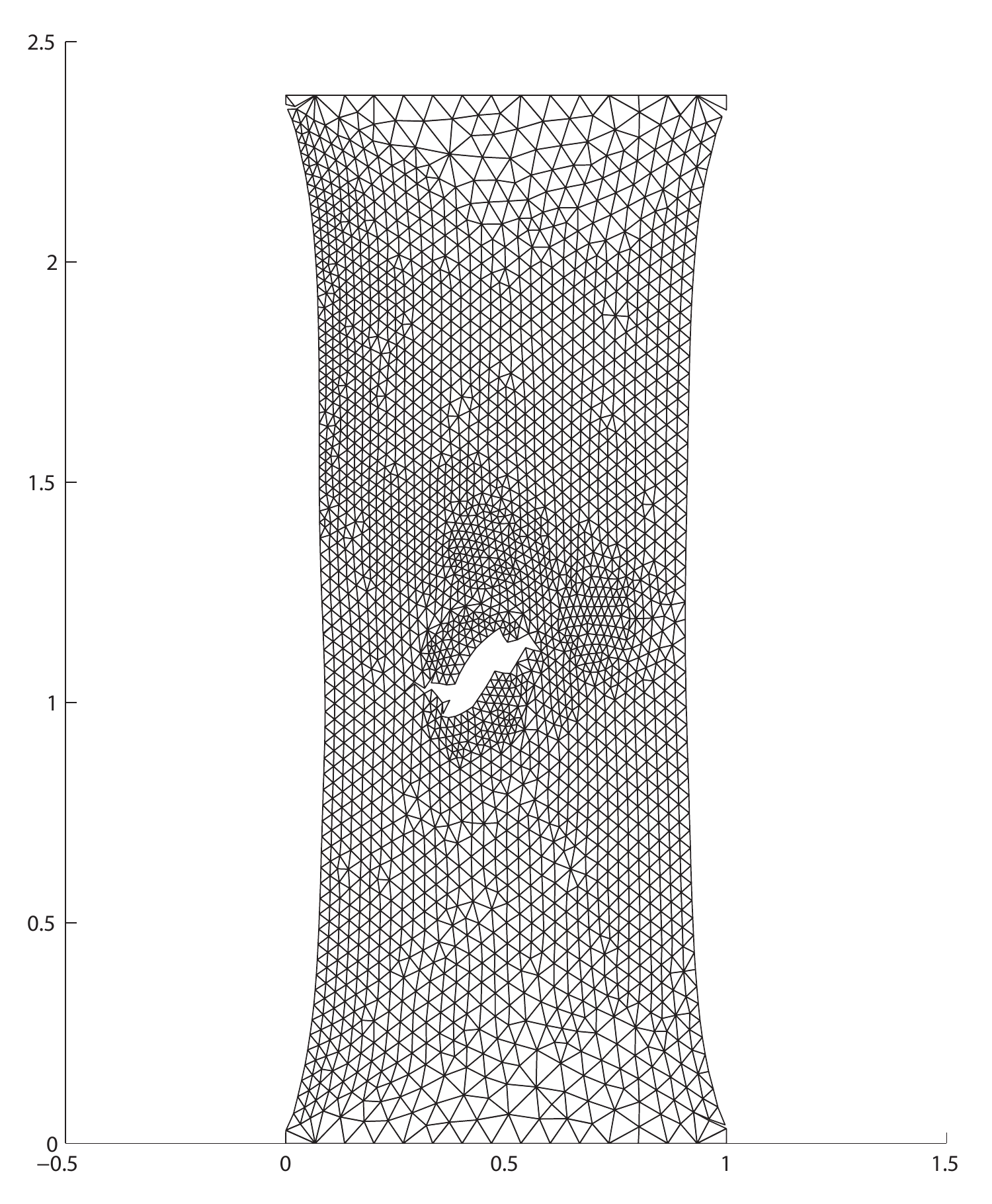}
		\includegraphics[scale=0.28]{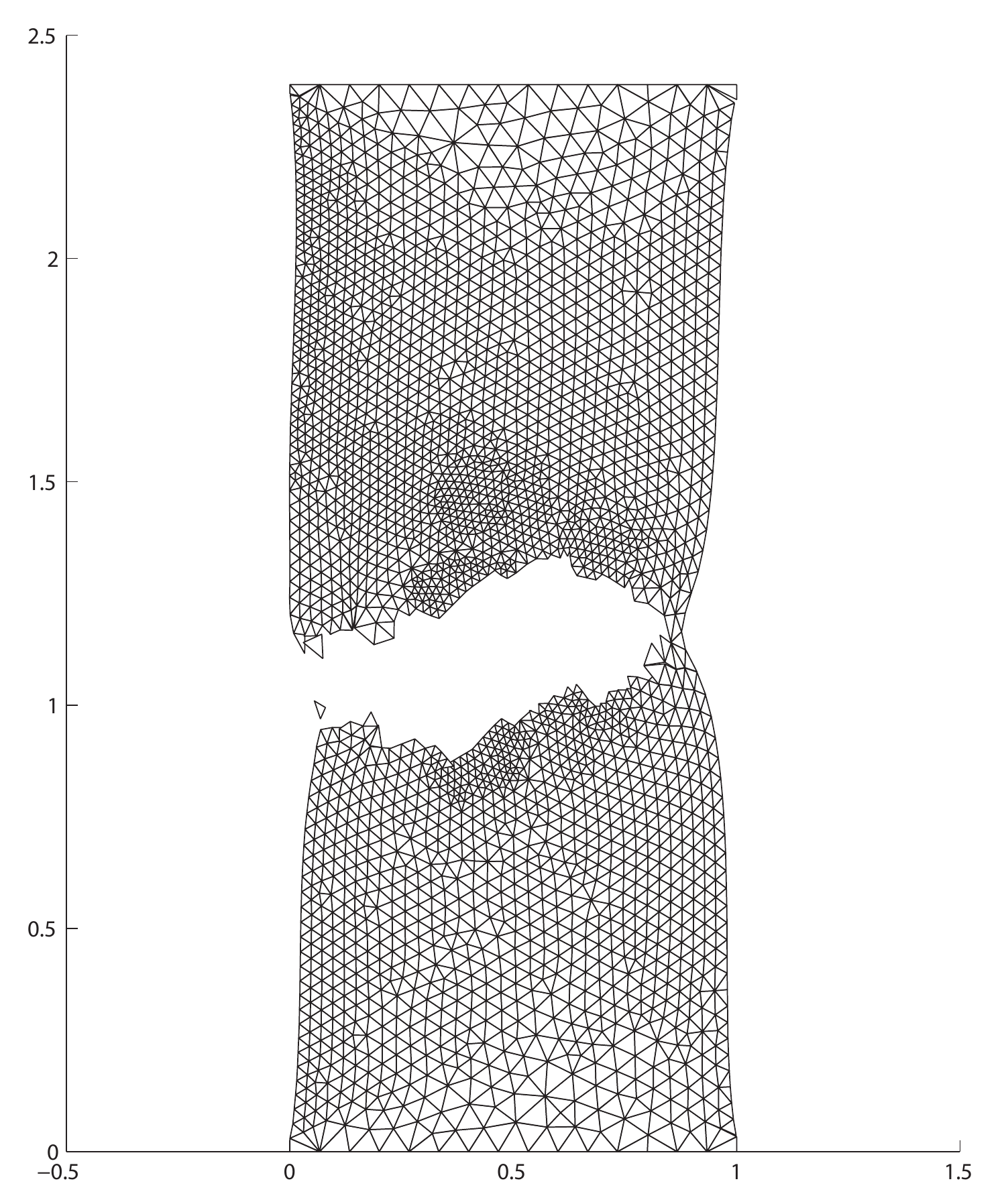}
	\caption{Increasing deformation from left to right for the first set-up.}
	\label{fig:SetUp1}
\end{figure}

For the first set-up, see Fig. \ref{fig:SetUp1}, the crack initiates as expected and then it propagates without considering the inclusions. In the second set-up, however, the crack is arrested by the stiffer inclusion boundary and deflects downwards around the lower right inclusion to finally to continue to propagate to the free edge of the specimen. It can be seen for both set-ups that there is virtually no compliance issues prior to any cracks forming. Both simulations, thus shows the applicability of the modeling technique. 

\begin{figure}[htbp]
	\centering
		\includegraphics[scale=0.28]{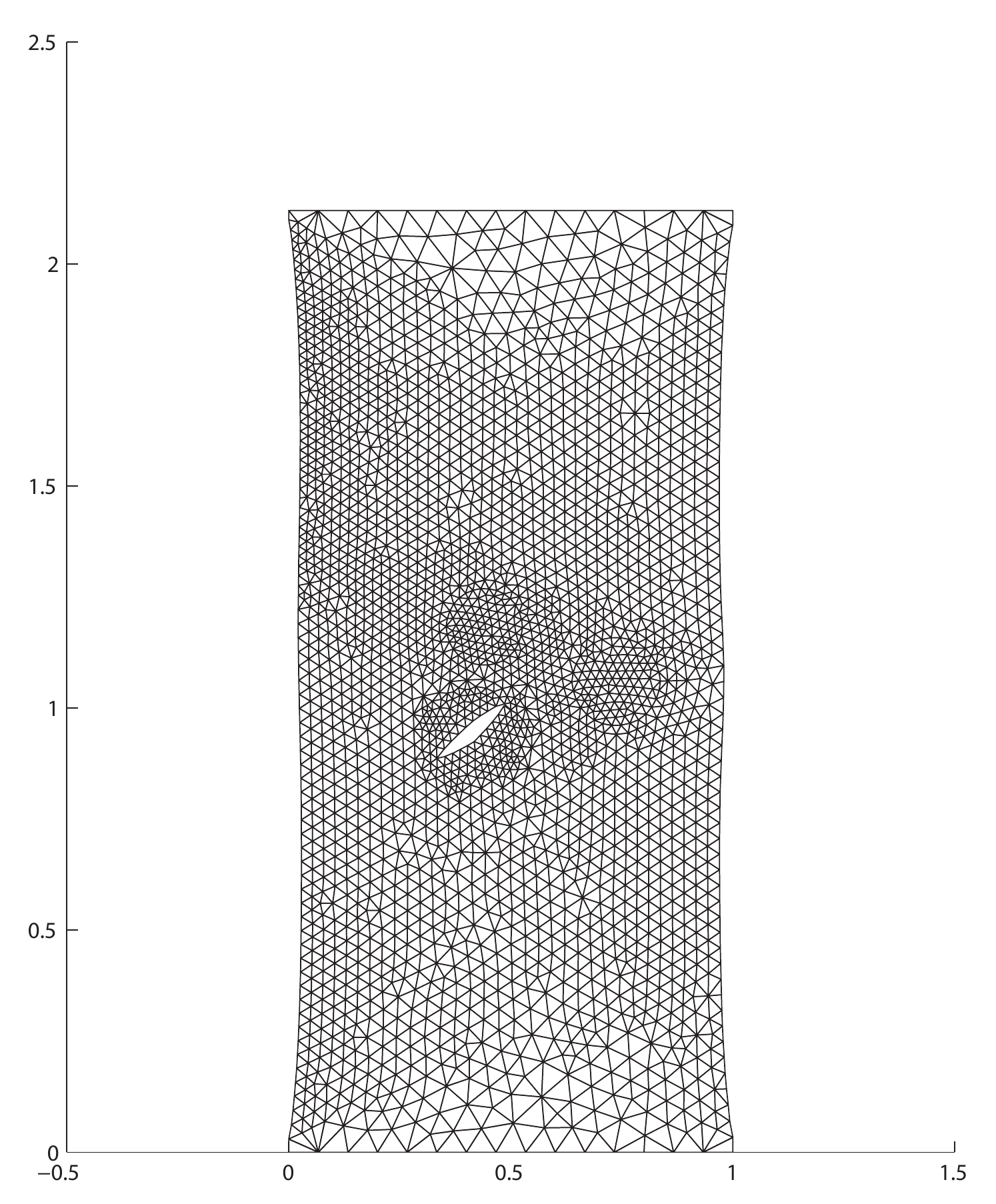}
		\includegraphics[scale=0.28]{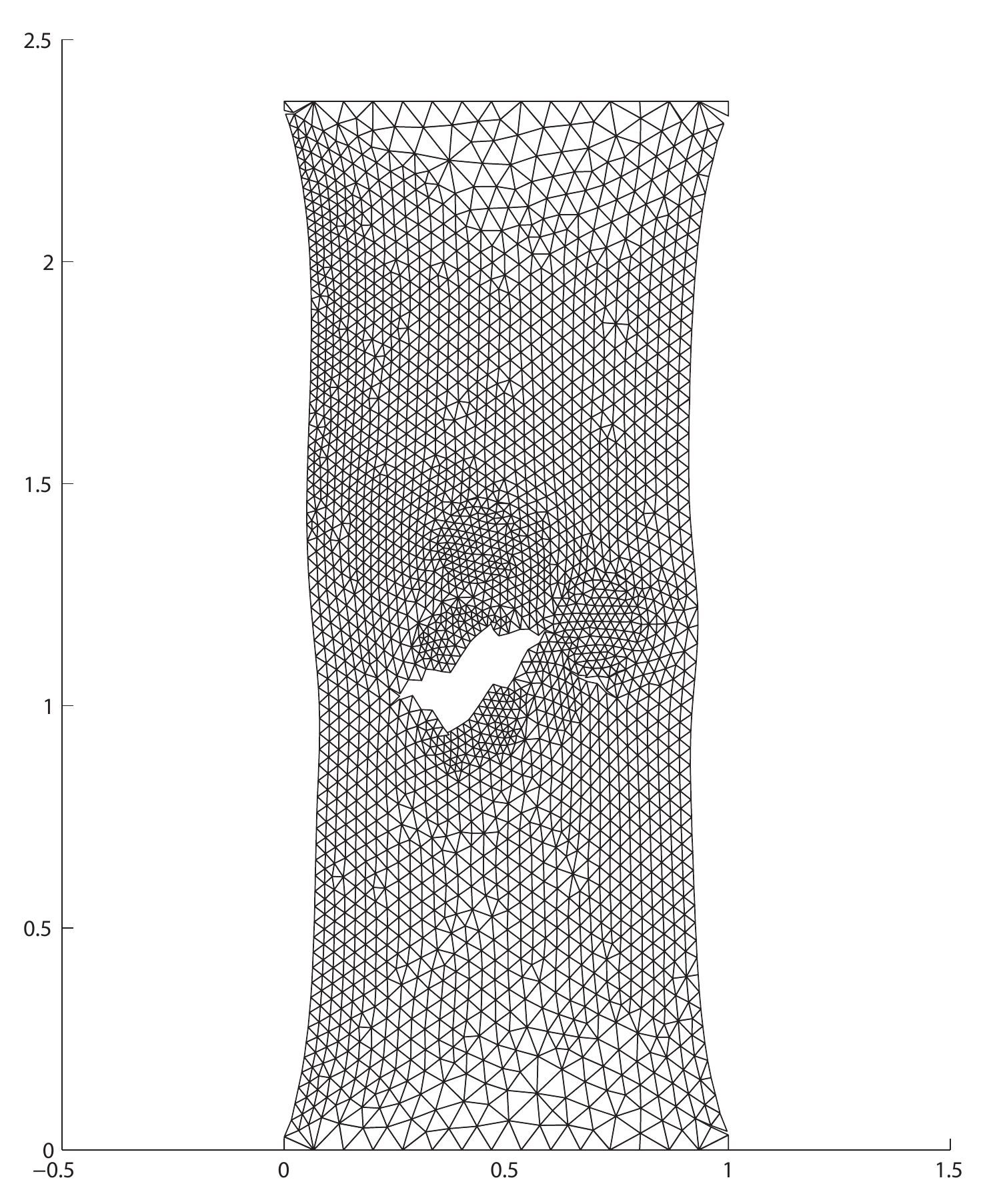}
		\includegraphics[scale=0.28]{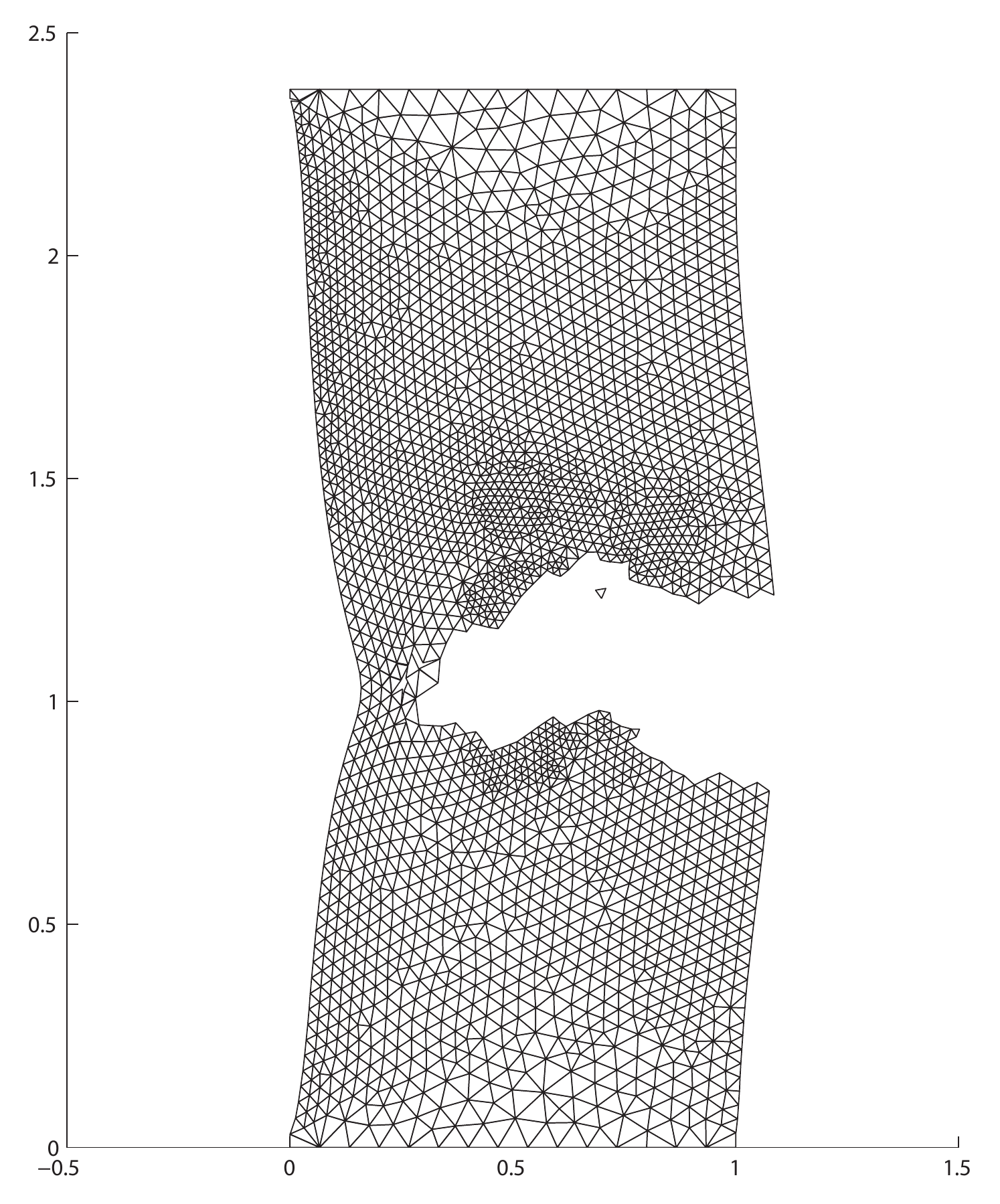}
	\caption{Increasing deformation from left to right for the second set-up.}
	\label{fig:SENcrack}
\end{figure}

\section{Concluding remarks}

In this paper, we have suggested an FE method which seamlessly blends the 
discontinuous Galerkin method with classical cohesive zone models. There is no need for interface elements
as the interelement stiffness is represented by a modification of the weak form. There is no need to identify
threshold values for transitions between discretization approaches since the same bilinear form is used for
all cases of interface stiffness. The method also directly allows for modeling cohesive zones between non--matching meshes, 
unlike the similar approach suggested previously in \cite{MeKuSt04}, which does not immediately generalize to this case.


\end{document}